\documentclass{article}
\usepackage{spconf,amsmath,graphicx}

\usepackage{enumitem}
\setlist{nosep, leftmargin=14pt}

\usepackage{mwe} 
\usepackage{boldline}
\usepackage{multirow}
\usepackage{multicol}
\usepackage{ulem}
\usepackage{color, soul}
\usepackage{amsfonts}
\usepackage[colorlinks,linkcolor=red]{hyperref}
\usepackage{url}

\title{Retinal Vessel Segmentation with Pixel-wise Adaptive Filters}
\name{
Mingxing Li$^{1*}$, Shenglong Zhou$^{1*}$, Chang Chen$^1$, Yueyi Zhang$^{1, 2\dag}$, Dong Liu$^1$, Zhiwei Xiong$^{1, 2}$
\vspace{-0.4cm}
} 
\address{$^1$ University of Science and Technology of China \\ $^2$ Institute of Artificial Intelligence, Hefei Comprehensive National Science Center}
\begin{document}
%
\maketitle
\begin{abstract}
\renewcommand{\thefootnote}{}
\footnotetext{$^*$ Equal contribution. $^\dag$ Corresponding author: zhyuey@ustc.edu.cn}

Accurate retinal vessel segmentation is challenging because of the complex texture of retinal vessels and low imaging contrast. Previous methods generally refine segmentation results by cascading multiple deep networks, which are time-consuming and inefficient. In this paper, we propose two novel methods to address these challenges. First, we devise a light-weight module, named multi-scale residual similarity gathering (MRSG), to generate pixel-wise adaptive filters (PA-Filters). Different from cascading multiple deep networks, only one PA-Filter layer can improve the segmentation results. Second, we introduce a response cue erasing (RCE) strategy to enhance the segmentation accuracy. Experimental results on the DRIVE, CHASE\_DB1, and STARE datasets demonstrate that our proposed method outperforms state-of-the-art methods while maintaining a compact structure. Code is available at \url{https://github.com/Limingxing00/Retinal-Vessel-Segmentation-ISBI2022}.
\end{abstract}
\vspace{-0.5em}
\section{Introduction}
\vspace{-0.5em}
\label{sec:intro}
Semantic segmentation is a fundamental task of biomedical image analysis, which can assist doctors in diagnosis and help biologists analyze cell morphology. In recent years, convolutional neural networks (CNNs) have shown remarkable effect on  biomedical image segmentation. Among them, U-Net \cite{ronneberger2015u} is the most widely used semantic segmentation network, which consists of an encoder to extract image features and a decoder to reconstruct the segmentation result. U-Net++ \cite{zhou2019unet++}  redesigns skip connections in the decoder, which improves the feature fusion and representation. 

For the retinal vessel segmentation, previous methods can be roughly divided into three categories. The first category  designs the topology-aware loss function to help the network recognize the critical structures \cite{hu2019topology,lan2020elastic}. The second category  utilizes multiple deep networks as the refinement module to refine the segmentation results \cite{wu2018multiscale,li2020iternet,li2020cascaded}. The third category enhances the capacity of the single network to obtain richer and more complex feature maps, such as those using the attention mechanism \cite{zhang2019attention,guo2020sa}. The method proposed in this paper belongs to the second category.
Although the second category has satisfactory results, 
the deep networks are time-consuming and inefficient.

To this end, we propose a method to utilize \textbf{only one layer} of pixel-wise adaptive filters (PA-Filters) to refine the segmentation results instead of using deep networks. In order to learn PA-Filters, we propose a light-weight module, named multi-scale residual similarity gathering (MRSG).
For each position on the initial segmentation map, MRSG generates a unique PA-Filter. Namely, unlike the traditional convolutional layer, the designed PA-Filters do not share weights to capture the texture of local regions better. 
Meanwhile, we propose a response cue erasing (RCE) strategy for further boosting the segmentation accuracy, which is implemented by an auxiliary branch. The RCE is responsible for erasing the corresponding position of the most confident pixels on the input image, depending on the output of the main branch.
We design a regularization loss to control the consistency of the dual branches, which makes the network more robust.
Experiments on three representative retinal vessel segmentation datasets (i.e. DRIVE, CHASE\_DB1 and STARE) validate that our efficient network achieves state-of-the-art performance.

\vspace{-0.7em}
\section{Method}
\vspace{-0.4em}
\subsection{Overview}
\begin{figure*}[t]
	\includegraphics[width=1\textwidth]{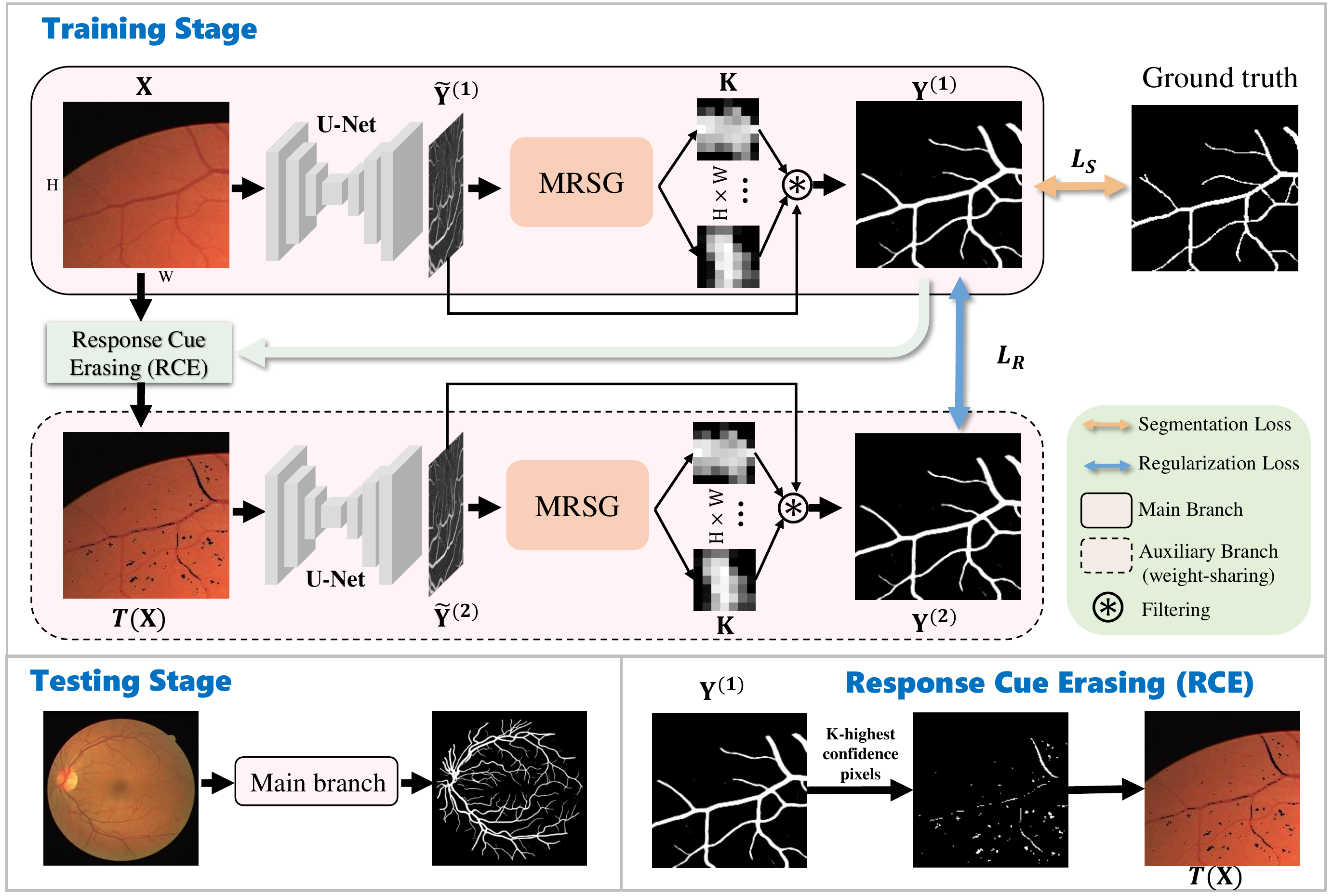} 
	\vspace{-1em}
	\caption{An overall framework of the proposed method. Two branches are adopted for the training stage with the patch-level input. Only the main branch is required for the testing stage with the entire input.}
	\vspace{-1em}
	\label{fig:framework}
\end{figure*}

As shown in Figure~\ref{fig:framework}, in the training stage, there are two branches in the network, the main branch and the auxiliary branch. The two branches are weight-sharing. The only difference is the input images of the auxiliary branch is processed via the RCE strategy. Take the main branch as an example, the input image, $\mathbf{X} \in \mathbb{R}^{3 \times H \times W}$, passes through a U-Net backbone to obtain a coarse segmentation map $\mathbf{\tilde{Y}}^{(i)}$ ($i=1, 2$). Then MRSG extracts the coarse segmentation map and input image to generate $H \times W$  PA-Filters $\mathbf{K}$ of size ${D \times D}$, where $D$ is a hyper-parameter.
Next, PA-Filters are applied to the corresponding local regions on the coarse segmentation map to obtain the final segmentation map $\mathbf{Y}^{(i)}$. During the testing stage, we only infer the main branch.

\vspace{-.7em}
\subsection{U-Net Backbone}
We adopt  U-Net as the backbone network $B$. 
Given $\mathbf{X}$ and $T(\mathbf{X})$, we can obtain the coarse segmentation map $\mathbf{\tilde{Y}}^{(i)} \in \mathbb{R}^{1 \times H \times W}$ ($i=1, 2$). $T(\cdot)$ denotes the RCE operation. The formulation of $\mathbf{\tilde{Y}}^{(i)}$ can be written as
\begin{equation}
\mathbf{\tilde{Y}}^{(1)} = B(\mathbf{X}, \theta); 
\mathbf{\tilde{Y}}^{(2)} = B(T(\mathbf{X}), \theta),
\end{equation}
where $\theta$ denotes parameters of U-Net. Here we set the channel number of the coarse segmentation map to 1 instead of one-hot encoding which is convenient for the following process. 

\begin{figure}[t]
	\centering
	\includegraphics[width=1\columnwidth]{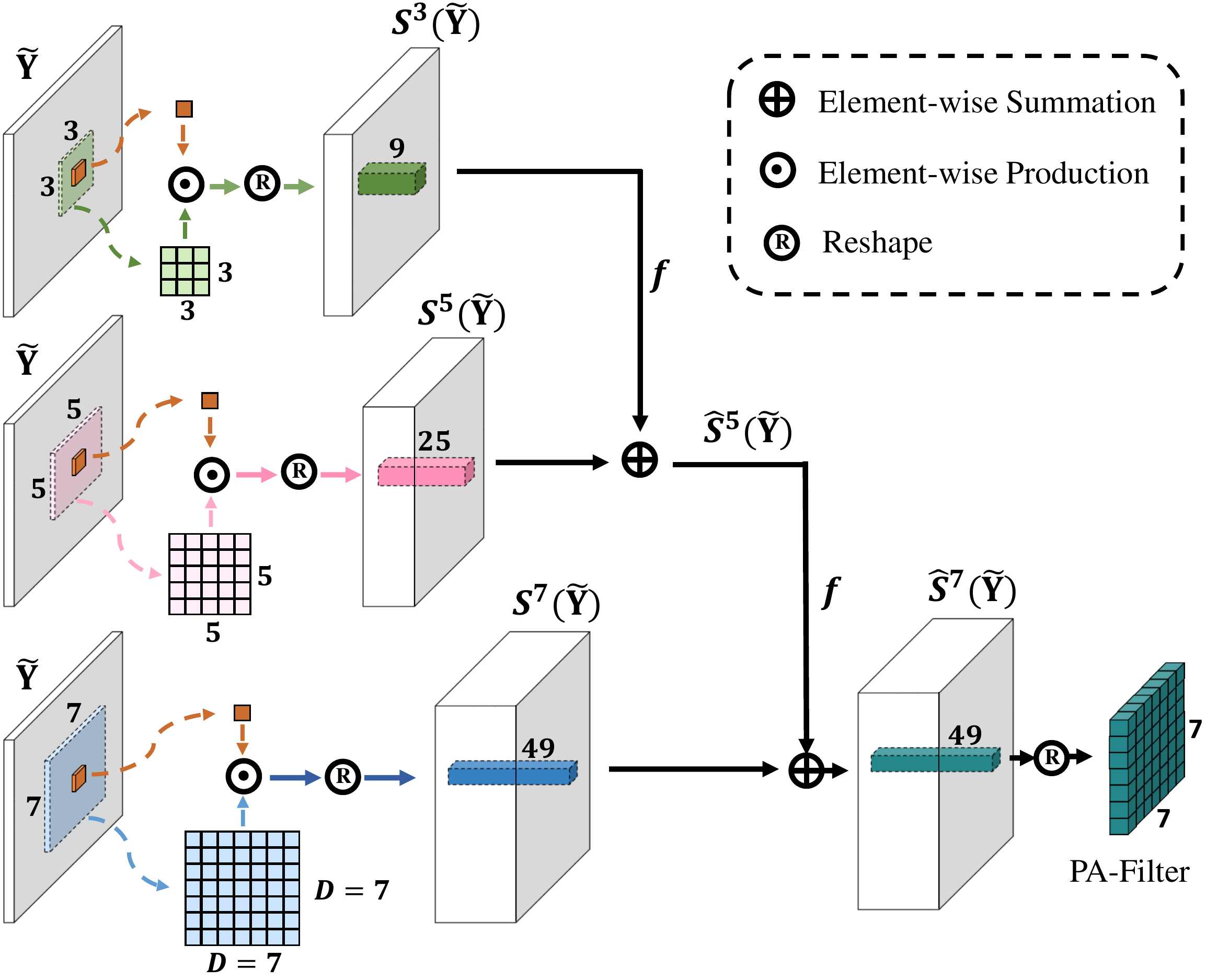} 
	\caption{The structure of the MRSG. Based on the coarse segmentation map ${\mathbf{\tilde{Y}}}$, the MRSG generates $H\times W$ PA-filters of size $D\times D$.}
	\vspace{-1em}
	\label{fig:MRSG}
\end{figure}

\subsection{Multi-scale Residual Similarity Gathering}
Inspired by prior works \cite{sun2018pwc,hui2018liteflownet}, we adopt the similarity volume for gathering the context information depending on the neighbour pixels. As shown in Figure~\ref{fig:MRSG}, for $\mathbf{\tilde{Y}}^{(i)}$, we calculate the similarity value $P^{\prime}_j$ by element multiplication between every pixel $P_{center}$ and its neighbouring of $d \times d$ pixels $P_j$ by the formula as follow:
\begin{equation}
P^{\prime}_j = P_j \times P_{center}
\end{equation}
where $j$ denotes the coordinate of the $d \times d$ region. Thus, for every pixel, we can obtain a local representation. Then we concatenate the local representation along the channel dimension to obtain the similarity volume ${S^d}(\mathbf{\tilde{Y}}^{(i)}) \in \mathbb{R}^{d^2 \times H \times W}$.
Furthermore, inspired by ACNet \cite{ding2019acnet} which indicates the skeletons are more important than the corners in a normal kernel, we find the closer pixels around the center pixel is more vital. Therefore, we propose a multi-scale residual scheme which adds the residual information for ${S^d}(\mathbf{\tilde{Y}}^{(i)})$ to obtain the final similarity volume $\hat{S^d}(\mathbf{\tilde{Y}}^{(i)})$. 
We make use of the similarity volume with a smaller $d$ for the residual information and introduce a bottleneck-style operation $f$ (a convolutional layer, a BatchNorm layer and a ReLU layer)  to sum up different volumes.
Based on the residual summation between similarity volumes, $\hat{S^D}(\mathbf{\tilde{Y}}^{(i)})$ can be constructed from $\{S^3(\mathbf{\tilde{Y}}^{(i)}), S^5(\mathbf{\tilde{Y}}^{(i)}), ..., S^D(\mathbf{\tilde{Y}}^{(i)})\}$ in a multi-scale procedure. We show the whole procedure and take $D=7$ as example in Equation~\ref{eq:f}:
\begin{equation}
\begin{aligned}
\hat{S^7}(\mathbf{\tilde{Y}^{(i)}}) &= S^7(\mathbf{\tilde{Y}^{(i)}}) + f(\hat{S^5}(\mathbf{\tilde{Y}^{(i)}})) 
\\
                     &= S^7(\mathbf{\mathbf{\tilde{Y}^{(i)}}}) + f(S^5(\mathbf{\tilde{Y}^{(i)}}) + f(S^3(\mathbf{\tilde{Y}^{(i)}}))).
\end{aligned}
\label{eq:f}
\end{equation}
After obtaining $\hat{S^D}(\mathbf{\tilde{Y}}^{(i)})$, we reshape $\hat{S^D}(\mathbf{\tilde{Y}}^{(i)}) \in \mathbb{R}^{D^2 \times H \times W}$ into $H\times W$ PA-Filters of size $D\times D$. 
Then PA-Filters are applied to the corresponding local regions on the coarse segmentation map to obtain the final segmentation map $\mathbf{Y}^{(i)}$.

\subsection{Response Cue Erasing}
To further exploit the potential of the network, we add an auxiliary branch and apply an RCE strategy for the input image of the auxiliary branch. As shown in Figure~\ref{fig:framework}, we adopt the RCE to generate erased regions and adopt the regularization loss to control the consistency of the dual branches.
The RCE has two steps. First, select the spatial position set $\{y^{(1)}_j\}$, $j \in [0, k-1]$, corresponding to the $k$ highest confidence pixels of the coarse segmentation map $\mathbf{\tilde{Y}}^{(1)}$, where both the foreground and background are considered. Second, erase the spatial position set $\{y^{(1)}_j\}$ of the input image.
Different from random erasing which cannot capture the structures, the RCE generates structure-dependent mask on the input image.

\subsection{The Overall Loss Function}
We choose the dice loss \cite{milletari2016v} which measures the difference between the label and the main branch output as the segmentation loss $L_{S}$. Besides, we propose the regularization loss $L_{R} = ||\mathbf{Y}^{(1)}- \mathbf{Y}^{(2)}||_2$ for the dual branches, which can constrain the consistency of the two outputs.
The overall loss $L$ is computed as $L = L_S+\lambda L_R.$



\section{Experiments and Analysis}
\subsection{Datasets}
We evaluate the proposed method on three popular retinal vessel segmentation datasets, DRIVE, CHASE\_DB1 and STARE. Specifically, DRIVE \cite{staal2004ridge} consists of 40 retinal images of size $565\times 584$ from a diabetic retinopathy screening program. Following the official partition, the training set has 20 images and the test set has the other 20 images.
CHASE\_DB1 \cite{owen2009measuring} contains 28 retinal images of size $999\times 960$.
STARE \cite{hoover2000locating} contains 20 retinal images of size $700\times 605$. We follow the setting of the method in \cite{li2020iternet}, which divides the first 20/16 images as the training set, and the last 8/4 images as the test set for these two datasets respectively.

\subsection{Implementation Details}
\label{subsec:detial}
In the experiments, we utilize Pytorch (version 1.1) to implement the proposed method. An NVIDIA GTX 1080Ti is used for training and testing.
During the training stage, we only use the flipping data augmentation. We minimize our loss using Adam, whose learning rate is 0.005 and fixed on the all datasets. We adopt the unified patch training strategy and set the patch size as 0.3 times the input image size. Thus the sampled patch sizes for the DRIVE, CHASE\_DB1 and STARE datasets are $169\times175$, $299\times288$ and $210\times181$ respectively.  We set batchsize 4 and maximum iteration 6000  on the three datasets. To balance the performance and computational burden, we choose $D = 5$ for the PA-Filters in our experiments.
We choose the suitable hyper-parameters $k$ and $\lambda$ according to different datasets.

\subsection{Quantitative and Qualitative Evaluation.}
\begin{figure}[t]
	\includegraphics[width=0.98\columnwidth]{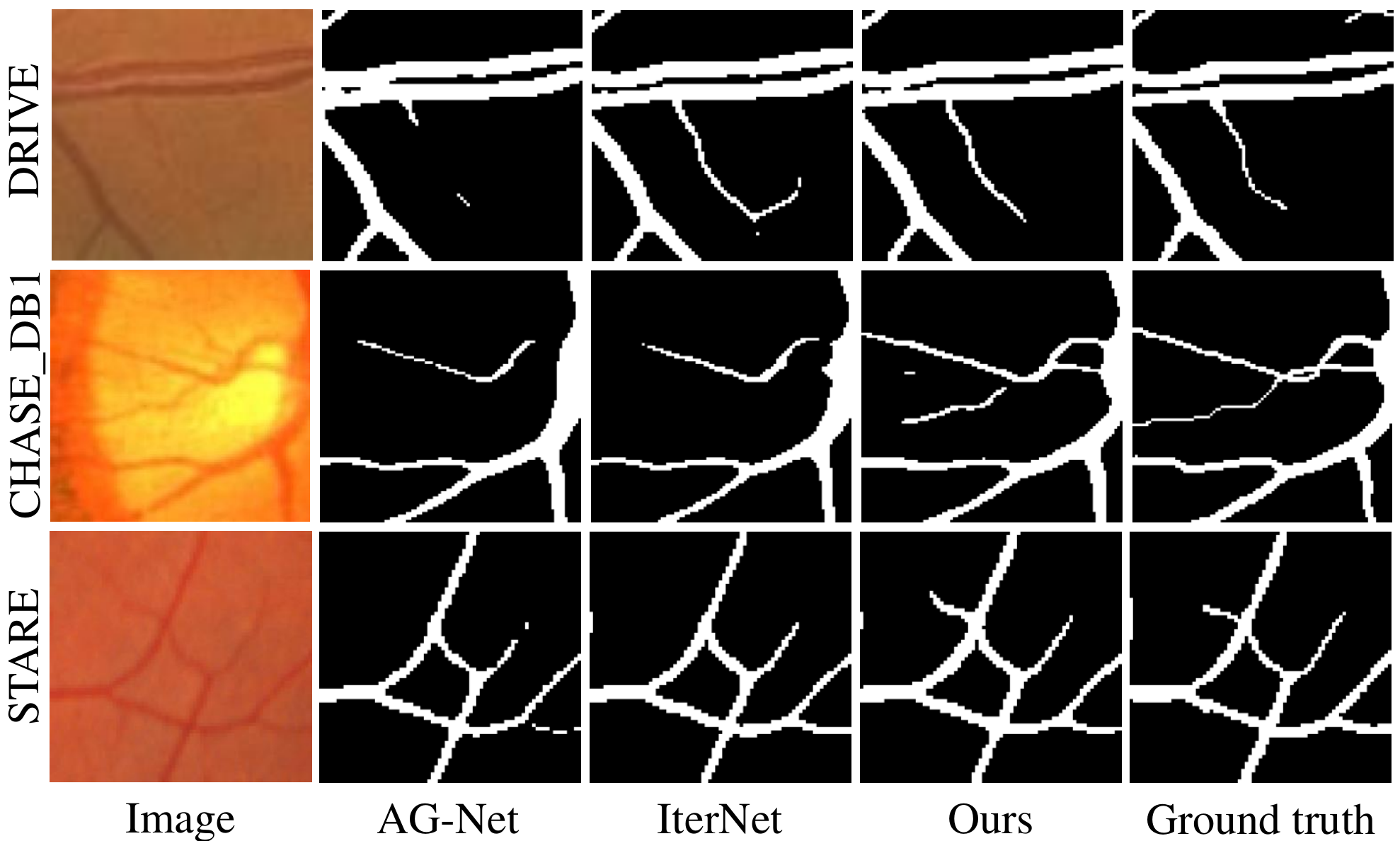} 
	\vspace{-0.8em}
	\caption{Visualized results on the DRIVE, CHASE\_DB1 and STARE datasets.}
	\label{fig:results}
	\vspace{-1em}
\end{figure}

\begin{table*}[]
\begin{center}
\setlength{\tabcolsep}{0.7em}
{\renewcommand{\arraystretch}{1.0}
\caption{Experimental results on the DRIVE,  CHASE\_DB1 and STARE datasets.   [Key: \textbf{Best}, \underline{Second} Best]}
\begin{tabular}{cc|ccc|ccc|ccc}
\hlineB{3}
\multirow{2}{*}{Method} & \multirow{2}{*}{Param.(MB)} & \multicolumn{3}{c|}{DRIVE} & \multicolumn{3}{c|}{CHASE\_DB1} & \multicolumn{3}{c}{STARE} \\ \cline{3-11} 
                         &          & F1       & AUC & ACC        & F1       & AUC & ACC    & F1     & AUC & ACC   \\ \hline
MS-NFN \cite{wu2018multiscale} & -   & -   & 98.07  & 95.67   & -  & 98.25  & 96.37 & -  & - & -   \\  
U-Net++ \cite{zhou2019unet++}  & 9.162  & 81.92  & 98.12  & 96.88  & 81.34  & 98.35  & 97.62  & 78.59  & 97.63 & 97.57 \\  
AG-Net \cite{zhang2019attention} & 9.330   & 80.79  & \uline{98.40}     & 96.87  & \uline{81.54} & \textbf{98.72} & \textbf{97.64} & 80.28   & 98.00 & 97.54 \\  
HR-Net \cite{wang2020deep} & 3.883  & \uline{82.50} & 98.20  & 96.93   & 81.22   & 98.30  & \uline{97.63} & 79.30  & 96.92 & 97.52  \\  
CTF-Net \cite{wang2020ctf} & -  & 82.41   & 97.88 & 95.67  & -  & -  & - & -  & - & -   \\
UCU-Net \cite{mishra2020data}& -  & -   & 97.24 & 95.40  & -  & 97.63 & 96.01 & -  & - & -   \\
IterNet \cite{li2020iternet} & 8.251  & \uline{82.50}     & 98.04  & 96.89  & 81.21  & 98.15  & 97.46 & \uline{81.33}  & 96.89 & \uline{97.82}   \\
SCS-Net \cite{wu2021scs} & 3.700 & - & 98.37 & \uline{96.97}  & - & \uline{98.67}  & 97.44 & - & \textbf{98.77} & 97.36   \\
Ours  & \textbf{2.013}  & \textbf{82.61} & \textbf{98.43} & \textbf{96.99}    & \textbf{81.67}    & 98.35    & 97.61 & \textbf{81.70}  & \uline{98.43} & \textbf{97.88}  \\  
\hlineB{3}
\end{tabular}
\label{tab:exp}
}
\end{center}
\vspace{-1.3em}
\end{table*}

We take F1-score (F1), area under curve (AUC), accuracy (ACC) as the metrics, which are  evaluated by the open source \cite{zhang2019attention}.  Table~\ref{tab:exp} summarizes the parameter (Param.) and metrics of each state-of-the-art (SOTA) method on the DRIVE, CHASE\_DB1 and STARE datasets. 
We can observe the proposed method has the best F1-score, surpassing other  SOTA  methods on all three datasets. Although AG-Net has the best AUC on the CHASE\_DB1 dataset, the parameter of the proposed method is $4 \times$ smaller than AG-Net, which shows the compactness of the proposed method. We also show the segmentation results on three datasets in Figure~\ref{fig:results}.  Compared with other SOTA  methods, our segmentation results have more detailed textures and complete structures. 

\begin{figure}[t]
	\includegraphics[width=0.98\columnwidth]{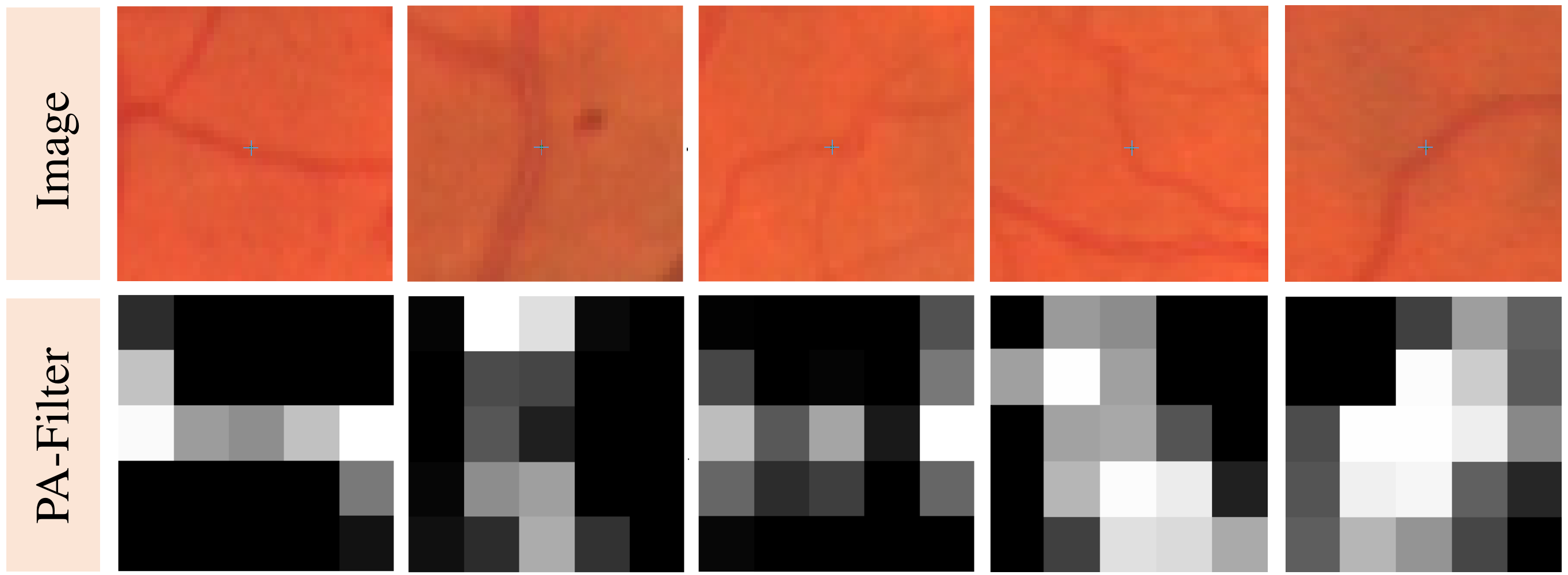} 
	\caption{The PA-Filters w.r.t the central points in the $41\times41$ region. The patches are sampled from the DRIVE dataset. White means high response, black means low response.}
	\label{fig:show_all}
\end{figure}

\begin{figure}[t]
	\includegraphics[width=0.98\columnwidth]{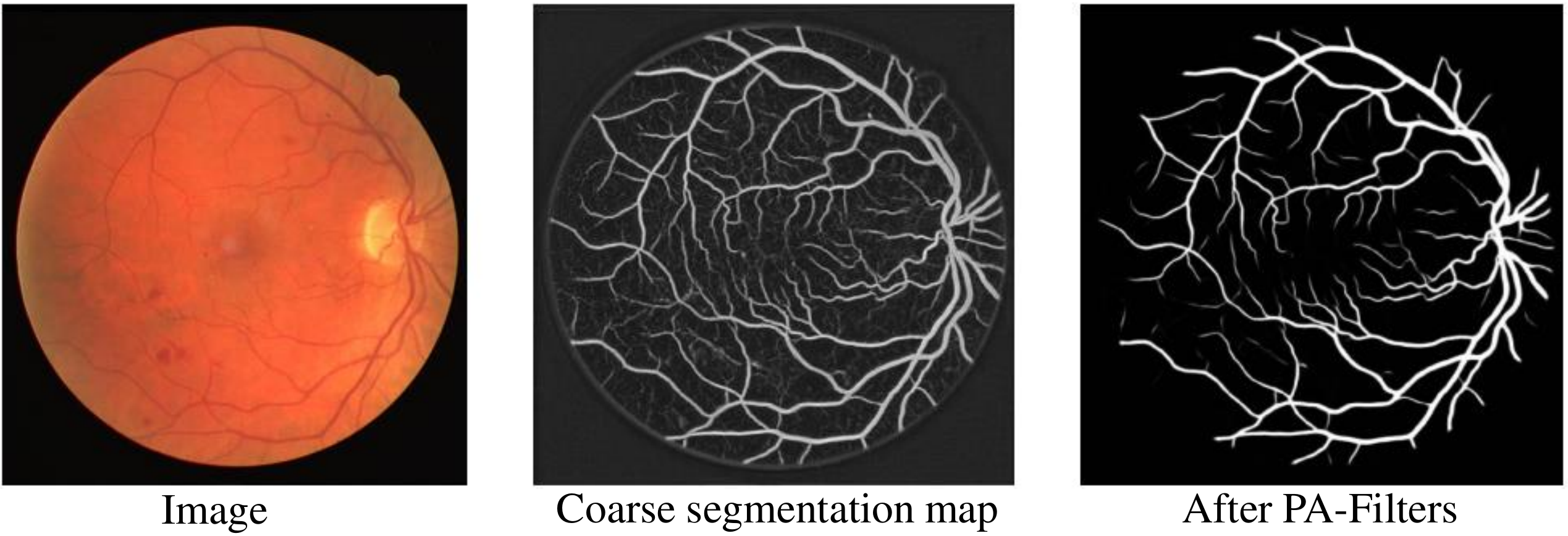} 
	\vspace{-1em}
	\caption{Visualized middle feature on the DRIVE dataset.}
	\label{fig:middle}
	\vspace{-1.2em}
\end{figure}
\subsection{Ablation Study}
To verify the contribution of each component in the proposed method, we conduct the ablation study. As shown in Table~\ref{tab:ablation}, we evaluate the effectiveness of the PA-Filters and the RCE strategy. When we choose PA-Filters of size $5\times 5$, the parameter of the network only increases by 0.012 MB, but F1-score increases by 3.1\%.   For the PA-Filters, we evaluate the impact of different kernel sizes without the RCE strategy. As shown in Table \ref{tab:kernel_ablation}, a larger D achieves  better performance at the cost of the requirement of larger GPU memory. Although $D=7/9$ has better performance, it exceeds the memory with the fixed setting (Section \ref{subsec:detial}) on CHASE\_DB1. For the sake of  uniformity, our experiments are based on $D=5$.

\begin{table}[t]
\centering
\caption{Ablation study on the STARE dataset.}
\label{tab:ablation}
\resizebox{\columnwidth}{!}{
\begin{tabular}{cc|cccc|c}
\hlineB{3}

PA-Filters & RCE & Param.(MB) &  F1    & AUC   & ACC    & Time(ms) \\ \hline
$\times$        & $\times$ & 2.001  & 78.60 & 96.83 & 97.50 &   5.1  \\ 
$\times$        & $\surd$  & 2.001  & 79.46 & 97.31 & 97.67 &   5.1  \\ 
$\surd$        & $\times$  & 2.013   & 81.13 & 97.81 & 97.76  & 9.7   \\ 
$\surd$        & $\surd$   & 2.013  & \textbf{81.70} & \textbf{98.43} & \textbf{97.88}  & 9.7  \\  \hlineB{3}
\end{tabular}}
\end{table}

\begin{table}[t]
\centering
\caption{Ablation of different kernel sizes of the PA-Filters w/o RCE on the STARE dataset.}
\label{tab:kernel_ablation}
\resizebox{\columnwidth}{!}{
\begin{tabular}{c|ccccc}
\hlineB{3}
Kernel Size (D)       & w/o PA-Filters & 3     & 5     & 7     & 9     \\ \hline
Param.   & 2.001             & 2.011 & 2.013 & 2.024 & 2.060  \\ 
Memory (GB) & 1.48            & 3.08 & 5.48 & 7.80  & 9.28  \\  
F1-score & 78.60             & 80.96 & 81.13 & 81.78 & \textbf{81.95} \\  
AUC      & 96.83             & 97.47 & \textbf{97.81} & 97.79 & 97.73 \\  \hlineB{3}

\end{tabular}
}
\end{table}

\subsection{Interpretability of the Proposed Method}
In the training stage, we have no supervision for the generation of the PA-Filters. As shown in Figure~\ref{fig:show_all}, PA-Filters learned at the central pixel implicitly reconstruct the texture of the retinal vessels instead of the local segmentation results. Taking the local patch of the first column of Figure~\ref{fig:show_all} as an example, the PA-Filter learned from the center point is similar to the stripe. Note that the center point is on the border of retinal vessels. The learned PA-Filters implicitly learn the textures, which makes the coarse segmentation map pay attention to the vessel boundary. Therefore, as shown in Figure \ref{fig:middle}, the PA-Filter can refine the coarse segmentation results using only one layer.

\section{Conclusion}
In this paper, we propose PA-Filters and RCE strategy for retinal vessel segmentation.  Specifically, we firstly utilize a U-Net backbone to obtain a coarse segmentation map, based on which the PA-Filters are generated. We devise an MRSG module to generate the PA-Filters for refinement.
Moreover, an RCE strategy is proposed to further improve the performance. Experimental results on three representative retinal vessel datasets (DRIVE, CHASE\_DB1 and STARE) demonstrate the superiority of the proposed method.

\section{COMPLIANCE WITH ETHICAL STANDARDS}
Ethical approval was not required as confirmed by the license attached with the open access data.

\section{Acknowledgment}
This work was supported in part by Anhui Provincial Natural Science Foundation Grant No. 1908085QF256 and University Synergy Innovation Program of Anhui Province No. GXXT-2019-025.
\normalem
\bibliographystyle{IEEEbib}
\bibliography{strings}

\end{document}